\documentclass[10pt,twocolumn,floatfix]{revtex4-1}

\usepackage{graphics,graphicx}
\usepackage{amsmath, amssymb}
\usepackage{multirow}
\usepackage{color}
\usepackage{verbatim}
\usepackage{hyperref}
\usepackage[normalem]{ulem}  
\usepackage{ulem}
\usepackage{color}
\usepackage{cancel}
\usepackage{mathtools}
\usepackage{epstopdf}

\renewcommand\sout{\bgroup \color{blue}\ULdepth=-.5ex \ULset}

\renewcommand\sout{\bgroup\color{blue} \ULdepth=-.5ex \ULset}
\def\slashchar#1{\setbox0=\hbox{$#1$}  
\dimen0=\wd0     
\setbox1=\hbox{/} \dimen1=\wd1  
\ifdim\dimen0>\dimen1   
\rlap{\hbox to \dimen0{\hfil/\hfil}} 
#1     
\else     
\rlap{\hbox to \dimen1{\hfil$#1$\hfil}} 
/      
\fi}

\newcommand{\dd}{\mathrm{d}}

\begin{document}

\title{Reconciling multi-messenger constraints with chiral symmetry restoration}

\author{Micha\l{} Marczenko}
\email{michal.marczenko@uwr.edu.pl}
\affiliation{Incubator of Scientific Excellence - Centre for Simulations of Superdense Fluids, University of Wroc\l{}aw, PL-50204 Wroc\l{}aw, Poland}
\author{Krzysztof Redlich}
\author{Chihiro Sasaki}
\affiliation{Institute of Theoretical Physics, University of Wroc\l{}aw, PL-50204 Wroc\l{}aw, Poland}
\begin{abstract}
    We analyze the recent astrophysical constraints in the context of a hadronic equation of state (EoS), in which the baryonic matter is subject to chiral symmetry restoration. We show that with such EoS it is possible to reconcile the modern constraints on the neutron star (NS) mass, radius, and tidal deformability (TD). We find that the softening of the EoS, required by the TD constraint of a canonical $1.4~M_\odot$ NS, followed by a subsequent stiffening, required by the $2~M_\odot$ constraint, is driven by the appearance of $\Delta$ matter due to partial restoration of chiral symmetry. Consequently, a purely hadronic EoS that accounts for the fundamental properties of quantum chromodynamics linked to the dynamical emergence of parity doubling with degenerate masses of nucleons and $\Delta$ resonances can be fully consistent with multi-messenger data. Therefore, with the present constraints on the EoS, the conclusion about the existence of the quark matter in the stellar core may still be premature.
\end{abstract}
 \maketitle 

\section{Introduction}
\label{sec:introduction}

The advancements of multi-messenger astronomy on different sources have led to remarkable improvements in constraining the equation of state (EoS) of dense, strongly interacting matter. The modern observatories for measuring masses and radii of compact objects, the gravitational wave interferometers of the LIGO-VIRGO Collaboration (LVC)~\citep{Abbott:2018exr, LIGOScientific:2018hze}, and the X-ray observatory Neutron Star Interior Composition Explorer (NICER) provide new powerful constraints on their mass-radius (M-R) profile~\citep{Riley:2019yda, Riley:2021pdl, Miller:2019cac, Miller:2021qha}. These stringent constraints allow for a detailed study of the neutron star (NS) properties and ultimately the microscopic properties of the EoS. In particular, the existence of $2~M_\odot$ NSs requires that the EoS must be stiff at intermediate to high densities to support them from gravitational collapse. At the same time, the tidal deformability (TD) constraint of a canonical $1.4~M_\odot$ NS from the GW170817 event implies that the EoS has to be fairly soft at intermediate densities. \cite{Fattoyev:2017jql} suggested that the softening of the EoS at intermediate densities required to comply with the TD constraint, together with the subsequent stiffening at high densities required to support $2~M_\odot$ NSs, may be indicative for a phase transition in the stellar core. In several works, this transition was associated with a possible occurrence of a hadron-quark phase transition and, thus, the presence of deconfined quark matter. This has been achieved by systematic analyses of recent astrophysical observations within simplistic approaches, such as the constant-speed-of-sound (CSS) model~\citep{Alford:2013aca, Alford:2017qgh, Li:2021sxb}. Although such schemes are instructive, they are not microscopic approaches. They provide interesting heuristic guidance, but cannot replace more realistic dynamical models for the EoS, which accounts for the fundamental properties of quantum chromodynamics (QCD), the theory of strong interactions, i.e., a self-consistent treatment of the chiral symmetry restoration in the baryonic sector. Recently, the possible phase transition to quark matter in the stellar core was also addressed in the supernova~\citep{Fischer:2017lag} and binary NS merger simulations~\citep{Bauswein:2018bma,Blacker:2020nlq,Bauswein:2020aag}.

Understanding the NS physics is at the interface with QCD. Masses and radii of pulsars can provide stringent constraints on the EoS and phase structure of QCD in a region of the QCD phase diagram that is inaccessible to terrestrial experiments and present techniques of lattice QCD (LQCD) simulations. At the same time, the recent LQCD results exhibit a clear manifestation of the parity doubling structure for the low-lying baryons around the chiral crossover~\citep{Aarts:2018glk}. The masses of the positive-parity ground states are found to be rather temperature-independent, while the masses of negative-parity states drop substantially when approaching the chiral crossover temperature $T_c$. The parity doublet states become almost degenerate with a finite mass in the vicinity of the chiral crossover. The observed behavior of parity partners is likely an imprint of the chiral symmetry restoration in the baryonic sector of QCD and is expected to occur also in cold dense matter. Such properties of the chiral partners can be described in the framework of the parity doublet model~\citep{Detar:1988kn, Jido:2001nt}. The model has been applied to hot and dense hadronic matter, and neutron stars (see, e.g,~\cite{Dexheimer:2007tn, Zschiesche:2006zj, Benic:2015pia, Marczenko:2017huu, Marczenko:2018jui, Marczenko:2019trv, Marczenko:2020jma, Marczenko:2020omo, Marczenko:2020wlc, Sasaki:2010bp, Yamazaki:2019tuo, Motohiro:2015taa, Minamikawa:2020jfj}).

In this work, we utilize the parity doublet model for nucleonic and $\Delta$ matter~\citep{Takeda:2017mrm} to investigate the implications on the structure of neutron stars in the light of the recent results from LIGO-VIRGO and NICER.

\section{Equation of State}
\label{sec:eos}

A hadronic parity doublet model~\citep{Takeda:2017mrm} is used to describe the NS properties. The model is based on the Lagrangian that contains mesonic and baryonic parity doublers (nucleon and $\Delta(1232)$) to form the proper chiral multiplets both in Nambu-Goldstone and Wigner phases. In this work, we consider a system with $N_f=2$. The baryonic degrees of freedom are coupled to the chiral fields $\left(\sigma, \boldsymbol\pi\right)$, the vector-isoscalar field ($\omega_\mu$), and the vector-isovector field ($\boldsymbol \rho_\mu$). The thermodynamic potential of the model in the mean-field approximation reads~\citep{Takeda:2017mrm}

\begin{equation}\label{eq:thermo_potential}
	\Omega = V_\sigma + V_\omega + V_\rho + \sum_{x=N,\Delta}\Omega_x\rm,
\end{equation}
where the index $x$ labels positive-parity and negative-parity spin-$1/2$ nucleons, i.e., $N\in \lbrace p,n;p^\star,n^\star \rbrace$, and spin-$3/2$ $\Delta$'s, i.e., \mbox{$\Delta \in \lbrace\Delta_{++,+,0,-};\Delta^\star_{++,+,0,-}\rbrace$}. Note that the negative-parity states are marked with the asterisk. The mean-field potentials in Eq.~\eqref{eq:thermo_potential} read

\begin{subequations}\label{eq:potentials}
\begin{align}
  V_\sigma &= -\frac{\lambda_2}{2}\sigma^2 + \frac{\lambda_4}{4}\sigma^4 - \frac{\lambda_6}{6}\sigma^6 - \epsilon\sigma \textrm,\label{eq:potentials_sigma}\\
  V_\omega &= -\frac{m_\omega^2 }{2}\omega^2\textrm,\label{eq:potentials_omega}\\
  V_\rho &= - \frac{m_\rho^2}{2}\rho^2\textrm,\label{eq:potentials_rho}
\end{align}
\end{subequations}
where $\lambda_2 = \lambda_4f_\pi^2 - \lambda_6f_\pi^4 - m_\pi^2$, and $\epsilon = m_\pi^2 f_\pi$. $m_\pi=140~$MeV, $m_\omega=783~$MeV, and $m_\rho=775~$MeV are the $\pi$, $\omega$, and $\rho$ meson masses, respectively, and $f_\pi=93~$MeV is  the pion decay constant. The kinetic part of the thermodynamic potential, $\Omega_x$, reads
\begin{equation}\label{eq:thermokin}
\Omega_x = \gamma_x \int \frac{\dd^3p }{(2\pi)^3} T \left(\ln\left(1-f_x\right) + \ln \left(1 - \bar f_x\right)\right)\rm,
\end{equation}
where the factors $\gamma_N = 2$ and $\gamma_\Delta=4$ denote the spin degeneracy of both parity partners for nucleons and $\Delta$'s, respectively. $f_x (\bar f_x)$ is the particle (antiparticle) Fermi-Dirac distribution function
\begin{subequations}
\begin{align}
	f_x &= \frac{1}{1+e^{\beta(E_x-\mu_x)}}\rm ,\\
	\bar f_x &= \frac{1}{1+e^{\beta(E_x+\mu_x)}}\rm,
\end{align}
\end{subequations}
with $\beta$ being the inverse temperature, the dispersion relation $E_x = \sqrt{\boldsymbol p^2+m_x^2}$ and $\mu_x$ is the effective chemical potential.

\begin{figure}[t!]\centering
  \includegraphics[width=.9\linewidth]{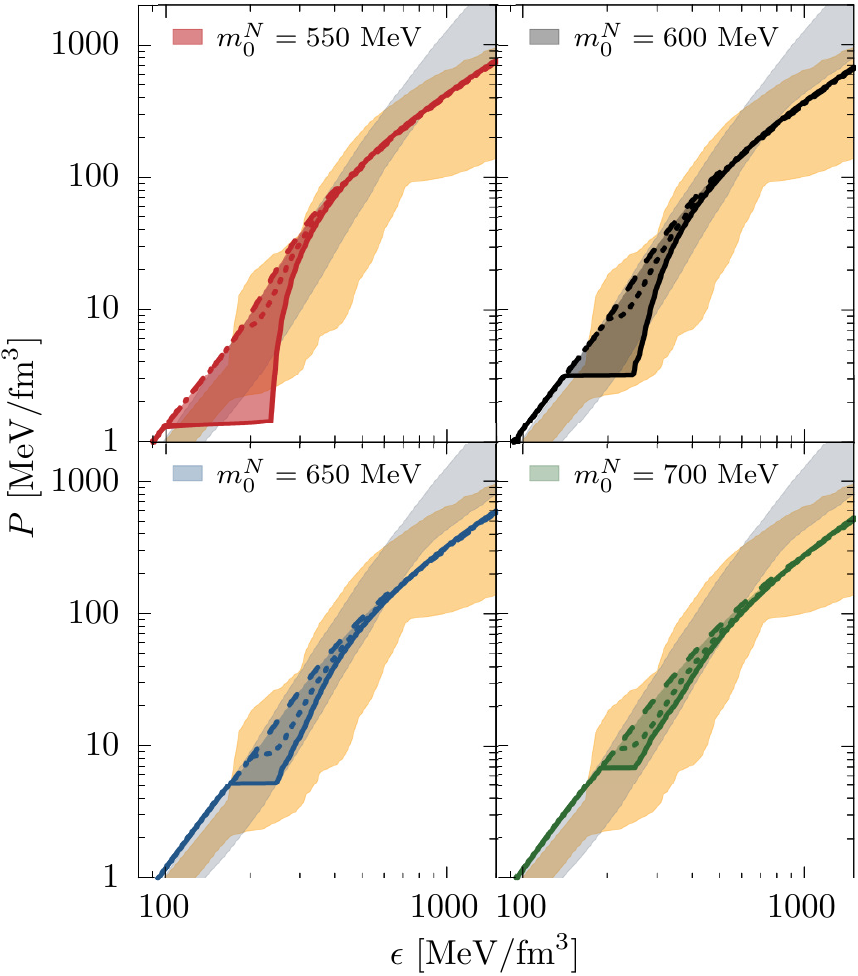}
  \caption{Thermodynamic pressure, $P$, under the NS conditions of $\beta$-equilibrium and charge neutrality, as a function of the energy density, $\epsilon$. The dashed lines correspond to the purely nucleonic EoSs. The solid lines correspond to the case $m_0^N=m_0^\Delta$. The region spanned between the two lines marks the results obtained for $m_0^N < m_0^\Delta$. The region enclosed by solid (dashed) and dotted lines show solutions where $\Delta$ matter enters the EoS through a first-order (crossover) transition. The orange- and grey-shaded regions show the constraints obtained by~\cite{Annala:2019puf} and~\cite{Abbott:2018exr}, respectively.}
  \label{fig:pressure}
\end{figure}

The masses of the positive- and negative-parity chiral partners are given by
\begin{equation}\label{eq:doublet_mass}
	m^x_\pm = \frac{1}{2}\left[\sqrt{\left(g_1^x+g_2^x\right)^2\sigma^2 + 4\left(m_0^x\right)^2} \mp \left(g^x_1-g^x_2\right)\sigma\right] \textrm,
\end{equation}
where $\pm$ sign denotes parity and $x=N,\Delta$. The spontaneous chiral symmetry breaking yields the mass splitting between the two baryonic parity partners in each parity doublet. When the symmetry is restored, the masses in each parity doublet become degenerate: $m_\pm^x(\sigma=0) = m_0^x$. The positive-parity nucleons are identified as the positively charged and neutral $N(938)$ states: proton ($p$) and neutron ($n$). Their negative-parity counterparts, denoted as $p^\star$ and $n^\star$, are identified as $N(1535)$ resonance~\citep{ParticleDataGroup:2020ssz}. The positive-parity $\Delta$ states are identified with $\Delta(1232)$ resonance. Their negative-parity chiral partners, $\Delta^\star$, are identified with $\Delta(1700)$ resonance~\citep{ParticleDataGroup:2020ssz}. For given chirally invariant mass, $m_0^x$, the parameters $g_1^x$ and $g_2^x$ are determined by the corresponding vacuum masses, $m_N = 939~$MeV, $m_{N^\star}=1500~$MeV, $m_\Delta = 1232~$MeV, $m_{\Delta^\star}=1700~$MeV.

The effective chemical potentials for nucleons and their chiral partners are given by
\begin{subequations}\label{eq:eff_chem_pot_N}
\begin{align}
	\mu_{p} &= \mu_{p^\star} = \mu_B + \mu_Q - g_\omega^N\omega - g_\rho^N\rho\rm,\\
	\mu_{n} &= \mu_{n^\star} = \mu_B - g_\omega^N\omega + g_\rho^N\rho\rm.
\end{align}
\end{subequations}
The effective chemical potentials for $\Delta$ and their chiral partners are given by
\begin{subequations}\label{eq:eff_chem_pot_D}
\begin{align}
	\mu_{\Delta_{++}} &= \mu_{\Delta^\star_{++}} = \mu_B + 2\mu_Q - g_\omega^\Delta\omega - 3g_\rho^\Delta\rho \rm,\\
	\mu_{\Delta_{+}} &= \mu_{\Delta^\star_{+}}  = \mu_B + \mu_Q - g_\omega^\Delta\omega - g_\rho^\Delta\rho\rm,\\
	\mu_{\Delta_{0}} &= \mu_{\Delta^\star_{0}}  = \mu_B - g_\omega^\Delta\omega + g_\rho^\Delta\rho\rm,\\
	\mu_{\Delta_{-}} &= \mu_{\Delta^\star_{-}}  = \mu_B -\mu_Q - g_\omega^\Delta\omega + 3g_\rho^\Delta\rho\rm.
\end{align}
\end{subequations}
The constants, $g_\omega^x$ and $g_\rho^x$ are the couplings of baryons to $\omega$ and $\rho$ mesons, respectively~\citep{Takeda:2017mrm}.

\begin{figure}[t!] \centering
    \includegraphics[width=.9\linewidth]{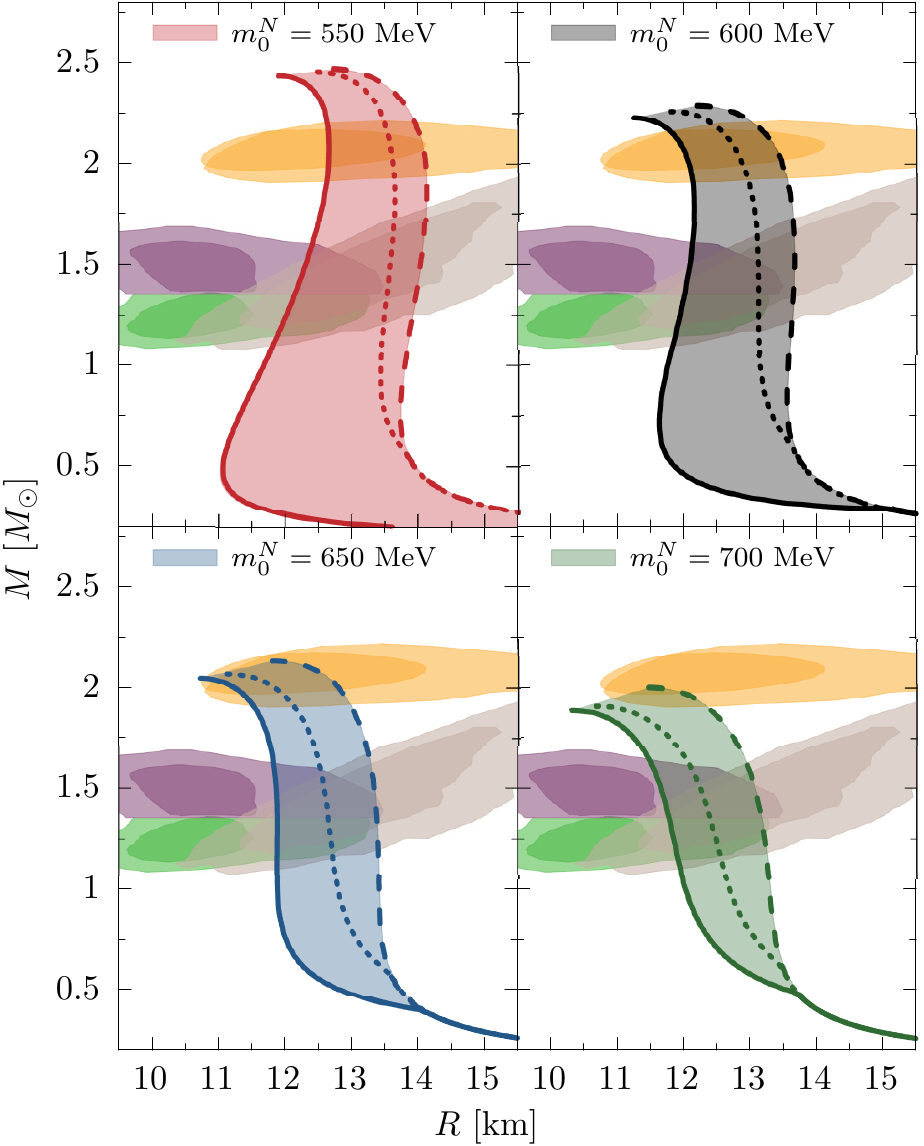}
    \caption{Calculated M-R sequences. The meaning of the lines and color coding is the same as in Fig.~\ref{fig:pressure}. The inner (outer) orange bands show the $1\sigma$ credibility regions from the NICER analysis of observations of the massive pulsar PSR J0740+6620 as dark orange~\citep{Riley:2021pdl} and light orange~\citep{Miller:2021qha}. The inner (outer) green and purple bands show 50\% (90\%) credibility regions obtained from the recent GW170817~\citep{Abbott:2018exr} event for the low- and high-mass posteriors. The inner (outer) gray regions correspond to the M-R constraint at 68.2\% (95.4\%) obtained for PSR J0030+0451 by the group analyzing NICER X-ray data~\citep{Miller:2021qha}.}
    \label{fig:m_r_band}
\end{figure}

The values of the $\lambda_4$, $\lambda_6$, and $g_\omega^N$ couplings are fixed by the properties of the nuclear ground state: the saturation density, $n_0=0.16~\rm fm^{-3}$, the binding energy, $\epsilon/n_B-m_N = -16~$MeV, and the compressibility, $K=240~$ MeV. The value of $g^N_\rho$ can be fixed by fitting the value of symmetry energy, $E_{\rm sym}=31~$MeV. The couplings of the $\Delta$ resonance to the meson fields are poorly constrained due to limited knowledge from experimental observations. The most advocated constraint was obtained by the analysis of electromagnetic excitations of the $\Delta$ baryon within the framework of the relativistic mean-field (RMF) model~\citep{Wehrberger:1989cd}. It puts a constraint on the relative strength of the scalar and vector couplings. Other phenomenological studies indicate an attractive $\Delta-N$ potential with no consensus on its actual size~\citep{Horikawa:1980cv, OConnell:1990njm, Lehr:1999zr, Nakamura:2009iq}. We note that in the parity doublet model the values of the nucleon-$\sigma$ and $\Delta-\sigma$ couplings, $g_1^x$ and $g_2^x$, are uniquely fixed by requiring the vacuum masses of the parity doublet states. On the other hand, the nature of the repulsive interaction among $\Delta$ resonances and their couplings to the $\omega$ and $\rho$ mean fields are still far from consensus. For simplicity, in the present study, we fix their values $g^\Delta_\omega=g^N_\omega$ and $g^\Delta_\rho=g^N_\rho$. We note that, in general, additional repulsion between $\Delta$'s would systematically shift their onset in the stellar sequence to higher densities. This eventually would prevent the neutron stars with $\Delta$ matter from existence in the gravitationally stable branch of the sequence. We note that this effect is similar as in the case of repulsive interactions between quarks~\citep{Marczenko:2020jma}.

In the present work, we take four representative values of $m_0^N=550$, $600$, $650$, $700~$MeV. Because the onset of $\Delta$ matter depends on the value of the chirally invariant mass $m_0^\Delta$~\citep{Takeda:2017mrm}, we systematically study the influence of $\Delta$ on the EoS and compliance with astrophysical constraints, i.e., $M_{\rm max} = (2.08 \pm 0.07)~M_\odot$~\citep{Fonseca:2021wxt}, as well as M-R and $\Lambda_{1.4} = 190^{+390}_{-120}$ from GW170817~\citep{Abbott:2018exr}. We note that the analysis of the data from GW170817 by~\cite{Abbott:2018exr} assumes a class of parametric EoSs without the possibility of a first-order phase transition. In principle, it is possible to parametrize an EoS so that it supports a phase transition. In this context, as shown in~\cite{Paschalidis:2017qmb}, such a class of EoSs that admits first-order phase transition is also consistent with the TD bounds from GW170817.

In this work, we put an additional constraint on the chirally invariant mass of $\Delta$. Namely, we require that $m_0^N \leq m_0^\Delta$. We chose this condition because too low values of $m_0^\Delta$ lead to the onset of $\Delta$ matter below the saturation density; thus, it spoils the properties of the ground state~\citep{Takeda:2017mrm}. The difference between $m_0^N$ and $m_0^\Delta$ can be attributed to the spin-spin interaction, similarly to hyper-fine splitting of the ground-state energy of the hydrogen atom in QED. We note that setting $m_0^\Delta = \infty$ suppresses the $\Delta$ states and the EoS effectively corresponds to the purely nucleonic EoS.

The obtained values of $L_{\rm sym}\approx82~$MeV at saturation for $m_0^N = 550-700~$MeV are in agreement with the commonly considered range of the parameter~\citep{Oertel:2016bki} and are found in other parity-doublet models~\citep{Motohiro:2015taa}. The most recent estimate, $L_{\rm sym} = 53^{+14}_{-15}~$MeV, is based on combined astrophysical data, PREX-II, and recent effective chiral field theory results~\citep{Essick:2021kjb}. Furthermore, the recent analysis within density functional theory yields $L_{\rm sym} = 54\pm8$~MeV~\citep{Reinhard:2021utv}. We note that $L_{\rm sym}$ does not depend on the choice of $m_0^\Delta$, due to the assumption that $\Delta$ matter appear at higher densities.

\begin{table}[t!]\begin{tabular}{|c||c|c|c|c|}
  \hline
  $m_0^N~$[MeV] & $R_{1.4}~$[km] & $\Lambda_{1.4}$ & $R_{\rm max}~$[km] & $M_{\rm max}~[M_\odot]$ \\ \hline\hline
  550           & 12.2,~14.0     & 506,~985        & 11.9,~ 12.7        & 2.44,~2.47 \\ \hline
  600           & 12.0,~13.7     & 394,~822        & 11.2,~ 12.2        & 2.23,~2.28 \\ \hline
  650           & 11.9,~13.4     & 321,~701        & 10.7,~ 11.8        & 2.05,~2.13 \\ \hline
  700           & 11.7,~13.1     & 275,~610        & 10.3,~ 11.5        & 1.89,~2.00 \\ \hline
  \end{tabular}
  \caption{Properties of canonical $1.4~M_\odot$ and the maximum-mass NSs. Values for $m_0^\Delta=m_0^N$ and purely nucleonic EoSs are separated by comma.}
  \label{tab:ns_params}
\end{table}

Fig.~\ref{fig:pressure} shows the calculated EoSs under the NS conditions of $\beta$-equilibrium and charge neutrality for selected values of $m_0^N$. The grey- and orange-shaded envelopes show the constraints derived by~\cite{Abbott:2018exr} and~\cite{Annala:2019puf}, respectively. To illustrate the effect of $\Delta$ matter on the EoS at intermediate densities, we show results obtained for purely nucleonic EoS (dashed line) together with the case $m_0^\Delta = m_0^N$ (solid line). The regions bounded by the two results correspond to the range spanned by solutions with $m_0^N < m_0^\Delta$ in each case. The region bounded by the solid and dotted lines corresponds to the range of EoSs with $\Delta$ appearing through a first-order transition. Consequently, the region between dotted and dashed lines shows the EoSs in which $\Delta$ matter appears smoothly. In general, the low-density behavior in each case is similar, until the deviations from the purely nucleonic EoSs are induced by the onset of $\Delta$ matter. The swift increase of the energy density is directly linked to the partial restoration of the chiral symmetry within the hadronic phase and resembled in the in-medium properties of dense matter in the parity doublet model. Most notably, it is associated with a drastic decrease of the negative-parity states in each parity doublet toward their asymptotic values, $m_0^x$. For instance, for $m_0^N=550~$MeV, the EoSs with $m_0^\Delta \approx m_0^N$ result in an appearance of $\Delta$ matter through a strong first-order transition, which results in a large jump of the energy density. Consequently, the EoS underestimates the pressure constraints at low densities. These constraints were derived under the assumption that there are no phase transitions in the EoS. It is reasonable to expect that models with more freedom may not fall within the pressure bounds obtained with more restrictive prior assumptions. Interestingly, the softening is followed by a subsequent stiffening, as compared to the purely nucleonic result, and the EoS reaches back the constraints at higher densities. This effect is more readily pronounced for smaller values of $m_0^\Delta$. For other parametrizations shown in the figure, the EoSs fall into the region derived by the constraint. We note that the notable change of the slope of $p(\epsilon)$ around $\epsilon\approx400-700~\rm MeV/fm^{3}$ was interpreted as evidence for the existence of quark matter in the NS cores~\citep{Annala:2017llu}. Interestingly, at higher densities, our results seem to follow this constraint and feature a similar change of the slope, regardless of the appearance of $\Delta$ matter. This behavior is linked to the restoration of chiral symmetry without additionally requiring the asymptotic pQCD behavior, as opposed to the interpretation given by~\cite{Annala:2017llu}. Thus, it does not necessarily signal the onset of deconfined quark matter in the NS cores.

\begin{figure}[t!]\centering
  \includegraphics[width=.9\linewidth]{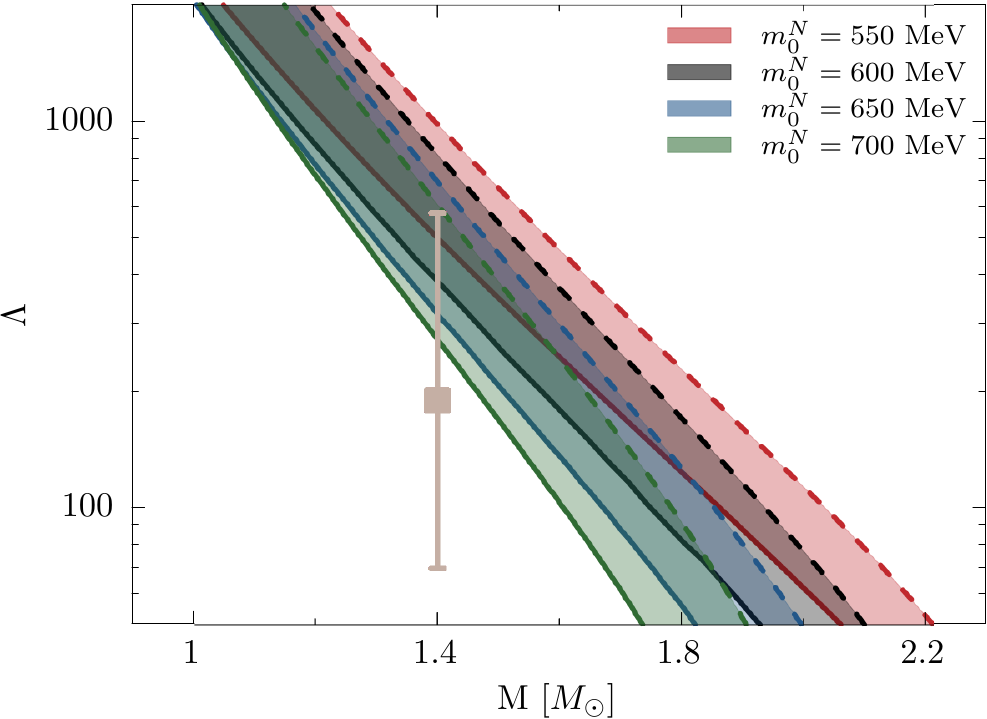}
  \caption{TD parameter $\Lambda$ as a function of neutron star mass. The error bar indicates the $\Lambda_{1.4} = 190^{+390}_{-120}$ constraint~\citep{LIGOScientific:2018hze}.}
  \label{fig:m_L_band}
\end{figure}

\section{Properties of neutron stars}
\label{sec:neutron_stars}

In Fig.~\ref{fig:m_r_band}, we show the M-R relations plotted up to the maximally stable solutions. Also shown are the state-of-the-art constraints: the high precision M-R analysis of the massive pulsar PSR~J0740+6620~\citep{Riley:2021pdl,  Miller:2021qha, Cromartie:2019kug, Fonseca:2021wxt} and PSR J0030+0451~\citep{Miller:2019cac} by the NICER collaboration, and the constraint from the recent GW170817 event~\citep{Abbott:2018exr}. The appearance of $\Delta$ matter affects the NS structure which is reflected in the M-R relations. As discussed in the previous section, the softening of the EoS due to the appearance of $\Delta$ is followed by a subsequent stiffening of the EoS. As a consequence, a substantial reduction of the radii of a $1.4~M_\odot$ NSs is observed (see Table~\ref{tab:ns_params}). In particular, the radii of the $1.4~M_\odot$ NS obtained in the purely nucleonic EoS reduce from $1.4~$km for $m_0^N=700~$MeV, to almost $2~$km for $m_0^N=600~$MeV when $m_0^\Delta=m_0^N$ case is considered. We note that the appearance of $\Delta$ matter was first connected with small NS radii by~\cite{Schurhoff:2010ph}. On the other hand, the decrease of the star's maximum mass is seen only mildly (see Table~\ref{tab:ns_params}). This, in turn, is a consequence of the subsequent stiffening of the EoS at higher densities, which allows for the massive stars to sustain from the gravitational collapse. \cite{Li:2018qaw} have shown that the earlier onset of $\Delta$ matter yields a larger maximum mass. We note that this stays in contrast to our work, where we find a mild decrease of the maximum mass. This is due to additional softening of the EoS at high density that is provided by the onset of the negative-parity $N^\star$ and $\Delta^\star$. This highlights the importance of chiral symmetry restoration in dense matter on the properties of the NS structure.

\begin{figure}[t!]
  \centering
  \includegraphics[width=.9\linewidth]{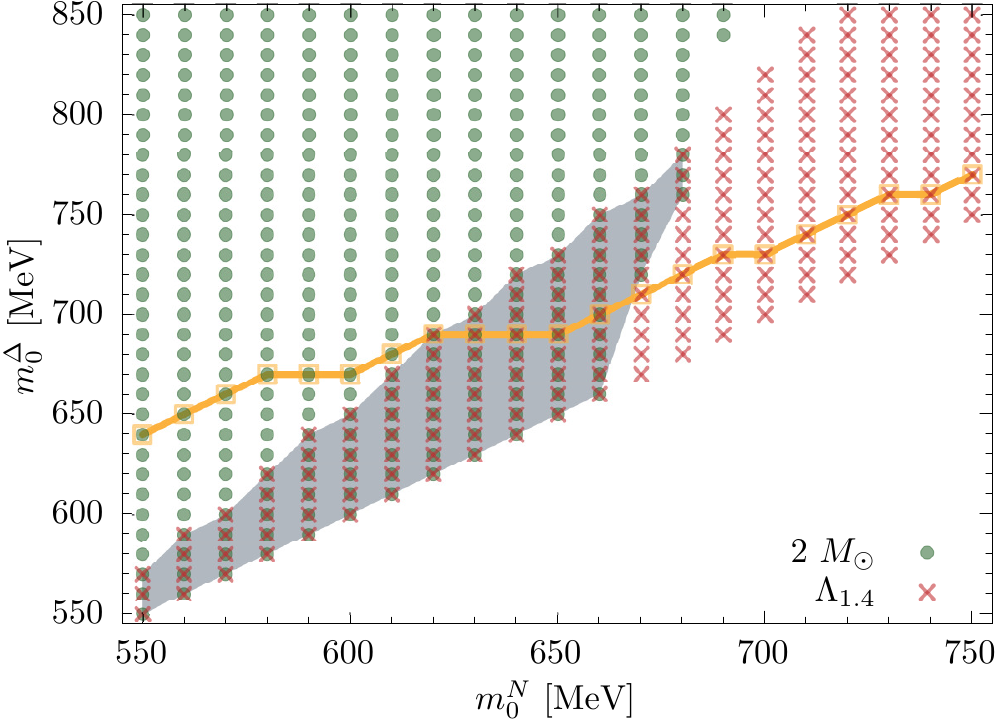}
  \caption{Allowed combinations of the model parameters, $m_0^N$ and $m_0^\Delta$. The green circles indicate configurations that fulfill the lower bound for the maximum mass constraint, $M_{\rm max} = (2.08\pm0.07)~M_\odot$~\citep{Fonseca:2021wxt}. The red crosses indicate configurations that are in accordance with the upper bound for the TD constraint, $\Lambda_{1.4} = 190^{+390}_{-120}$~\citep{Abbott:2018exr}. The gray-shaded area shows the region where the two constraints are fulfilled simultaneously. The orange points show configurations with the largest value of $m_0^\Delta$ for which the $\Delta$ matter appears through the first-order transition.}
  \label{fig:m0_m0}
\end{figure}

In Fig.~\ref{fig:m_L_band}, we show the dimensionless TD parameter $\Lambda$ as a function of the NS mass:

\begin{equation}
   \Lambda = \frac{2}{3} k_2 C^{-5} \textrm,
\end{equation}
where $C = M/R$ is the star compactness parameter. The parameter $\Lambda$ can be computed through its relation to the Love number $k_2$~\citep{Hinderer:2007mb}. For the purely nucleonic case, higher values of $m_0^N$ result in smaller values of TD. We note that even for $m_0^N=700~$MeV, the purely nucleonic EoS is too stiff to comply with the constraint. Recently, \cite{Reed:2021nqk} have shown that the stiffness of the EoS at intermediate densities is connected with its nuclear characteristics at saturation. Specifically, it is observed as a correlation between the value of TD of a $1.4~M_\odot$ NS and the value of $L_{\rm sym}$ at saturation. However, this conclusion is drawn upon the assumption that the EoS consists of purely nucleonic matter. Thus, the stiffness of the EoS due to large values of $L_{\rm sym}$ obtained in the parity doublet model creates tension with the TD constraint. The tension is commonly resolved with an onset of additional degrees of freedom at intermediate densities due to a strong first-order phase transition to quark matter~\citep{Li:2021sxb, Christian:2021uhd}. Such a conclusion about the existence of quark matter in the stellar core may, however, be premature. As we demonstrate in this work, strong phase transitions with a large latent heat may occur within hadronic matter due to the onset of $\Delta$ matter. This is seen in Fig.~\ref{fig:m_L_band}. If $\Delta$ matter is allowed in the EoS, TD reduces substantially, providing better agreement with the constraint. In general, smaller values of $m_0^\Delta$, i.e., earlier onset of $\Delta$ matter, result in smaller TDs of a canonical $1.4~M_\odot$ NS. The most significant reduction of TD arises for $m_0^\Delta = m_0^N$ (see Table~\ref{tab:ns_params}). For a given value of $m_0^N$, the value of $\Lambda_{1.4}$ can be reduced significantly, while keeping the same value of the slope of the symmetry energy when $\Delta$ matter is considered. We note that the TD parameter requires sufficiently soft EoS at intermediate densities. This is seen in the figure, where the best agreement with the constraint is obtained for the smallest values of $m_0^\Delta$. On the other hand, the $2~M_\odot$ requires a sufficiently stiff equation of state at higher densities. Inversely to TD, the most massive stars are obtained for the stiffest EoSs. Therefore, our results are following the TD and $2~M_\odot$ constraints.

Lastly, we discuss the allowed parameter space. Fig.~\ref{fig:m0_m0} shows the values of the chirally invariant masses, $m_0^N$ and $m_0^\Delta$ for which the TD and $2~M_\odot$ constraints are met. The TD constraint puts an upper limit on $m_0^\Delta$ and the purely nucleonic EoSs are ruled out. Therefore, the presence of $\Delta$ matter at intermediate densities is essential to comply with the TD constraint. The $2~M_\odot$ simultaneously constrains $m_0^N$ from above and puts a lower limit on $m_0^\Delta$. For sufficiently small $m_0^\Delta$, the softening provided by the early onset of $\Delta$ matter is enough to satisfy both constraints. We conclude that the chirally invariant masses cannot be significantly different from each other and can be estimated to lie in the range from $550$ to $680$~MeV. The orange squares (connected with a line for readability) denote configurations with the largest value of $m_0^\Delta$ for which the $\Delta$ matter appears through a first-order transition. Configurations above the squares correspond to the onset via a smooth crossover transition.

Based on the above results, we conclude that the possibility of a smooth appearance of $\Delta$ matter is consistent with the astrophysical constraints and cannot be excluded at the moment. Nevertheless, the swift softening of the EoS at low densities is caused by the partial restoration of chiral symmetry whose remnants are expected to be apparent in the in-medium properties of cold baryonic matter even in the absence of a sharp first-order phase transition. Therefore, as far as the present multi-messenger constraints are concerned, the observed properties of NSs can be explained by an EoS that accounts for the dynamical restoration of chiral symmetry, without any assumption about the existence of a quark matter core.

\section{Conclusion}
\label{sec:summary}

In this work, we have shown that it is possible to reconcile the current constraints from multi-messenger measurements within a purely hadronic EoS, which accounts for the self-consistent treatment of the chiral symmetry restoration in the baryonic sector. To this end, we have adopted the parity doublet model for nucleonic matter including $\Delta(1232)$ resonance being subject to chiral symmetry restoration.

We have analyzed the properties of neutron stars and found that the scenario of the softening of the EoS at intermediate densities with the subsequent stiffening at high densities required by the $2~M_\odot$ and TD constraints can be realized within a hadronic EoS, where the softening of the EoS is obtained by an early onset of $\Delta$ matter due to partial restoration of chiral symmetry in the hadronic phase. Consequently, the radius of a canonical $1.4~M_\odot$ NS reduces substantially providing better agreement with the TD constraint from the GW170817 event, leaving the maximum mass in agreement with the $2~M_\odot$ constraint. Moreover, as we have demonstrated in this work, these characteristics do not necessarily imply the existence of a hadron-quark phase transition as proposed in recent studies, e.g.,~\citep{Annala:2019puf, Li:2021sxb, Christian:2021uhd}. 

From these studies, we conclude that due to the anticipated near-future advances in multi-messenger astronomy, it will become inevitable to link the observed macroscopic properties of NSs and their mergers to fundamental properties of strong interactions described by QCD, including chiral symmetry restoration. Furthermore, as proposed by~\cite{Bauswein:2018bma, Blacker:2020nlq, Bauswein:2020aag} in the context of a phase transition to quark matter, it would be interesting to verify if signatures of chiral symmetry restoration related to a crossover or first-order phase transition in the hadronic phase have an observable imprint on the GW emission from NS mergers, and, thus, can be measurable in future GW detections.

The constraints on the EoS, and in particular the chirally invariant masses, derived from multi-messenger astronomy can be further verified by the forthcoming large-scale nuclear experiments FAIR at GSI and NICA in Dubna, whose focus will be on the role of chiral symmetry restoration in the high-density nuclear matter.

\section*{Acknowledgements}
This work was partly supported by the Polish National Science Centre (NCN) under OPUS Grant No. 2018/31/B/ST2/01663 (K.R. and C.S.), Preludium Grant No. 2017/27/N/ST2/01973 (M.M.), and the program Excellence Initiative–Research University of the University of Wroclaw of the Ministry of Education and Science (M.M). K.R. also acknowledges the support of the Polish Ministry of Science and Higher Education. We acknowledge helpful comments from David Alvarez-Castillo, David Blaschke, and Armen Sedrakian.

\bibliography{biblio}

\end{document}